\documentclass[pra,aps,twocolumn,superscriptaddress,longbibliography]{revtex4-1}
\usepackage[colorlinks=true, citecolor=red, urlcolor=blue ]{hyperref}
\usepackage{graphicx}
\usepackage{bm}
\usepackage{amsmath,amsfonts}
\usepackage{amsthm}
\usepackage{hyperref}
\usepackage{xcolor}
\hypersetup{
    urlcolor=magenta,
    citecolor=blue
           }

\usepackage{braket}
\theoremstyle{plain}
\usepackage{color}
\usepackage{amssymb}
\usepackage{amsthm}
\usepackage{amsfonts}
\usepackage{float}
\usepackage{tabularx}
\usepackage{graphicx}
\usepackage[utf8]{inputenc}

\usepackage{mathtools}
\usepackage{esvect}
\usepackage{wrapfig}
\usepackage{amsthm}
\usepackage{verbatim}
\usepackage{bbm}
\usepackage[normalem]{ulem}

\usepackage{enumitem}
\usepackage{fmtcount}
\usepackage{booktabs}
\usepackage{csquotes}
\usepackage{epsfig}

\usepackage{tabularx}
\usepackage{graphicx}
\usepackage{amsmath}
\usepackage{braket}
\usepackage{latexsym}
\usepackage{bm}
\usepackage{graphics,epstopdf}
\usepackage{enumitem}
\usepackage{fmtcount}
\usepackage{booktabs}
\usepackage{csquotes}
\usepackage{epsfig}

\theoremstyle{plain}

\def\bea{\begin{eqnarray}}
\def\eea{\end{eqnarray}}
\def\ba{\begin{array}}
\def\ea{\end{array}}

\def\beq{\begin{equation}}
\def\eeq{\end{equation}}
\DeclareMathOperator*{\ot}{\otimes}

\usepackage[normalem]{ulem}
\usepackage{float}
\usepackage{graphicx}  
\usepackage{dcolumn}          

\usepackage{orcidlink}
\usepackage{amssymb}
\usepackage{appendix}
\usepackage{physics}   
\usepackage{mathtools}
\usepackage{esvect}
\usepackage{wrapfig}
\usepackage{amsthm}
\usepackage{verbatim}
\usepackage{bbm}

\usepackage[mathscr]{euscript}
\def\Tr{\operatorname{Tr}}

\def\({\left(}
\def\){\right)}
\def\[{\left[}
\def\]{\right]}

\begin{document}

\title{Effects of the detection loophole on rival entanglement attestation techniques}

\author{Paranjoy Chaki\,\orcidlink{0009-0000-5693-5516}}
\affiliation{Harish-Chandra Research Institute,  A CI of Homi Bhabha National
Institute, Chhatnag Road, Jhunsi, Prayagraj - 211019, India}

\author{Kornikar Sen\,\orcidlink{0000-0002-7007-0843}}
\affiliation{Departamento de Física Teórica, Universidad Complutense, 28040 Madrid, Spain}
\affiliation{Harish-Chandra Research Institute,  A CI of Homi Bhabha National
Institute, Chhatnag Road, Jhunsi, Prayagraj - 211019, India}
\author{Ujjwal Sen\,\orcidlink{0000-0002-0091-5847}}
\affiliation{Harish-Chandra Research Institute,  A CI of Homi Bhabha National
Institute, Chhatnag Road, Jhunsi, Prayagraj - 211019, India}


\begin{abstract}

Entanglement detection is of crucial significance to quantum technologies and yet remains a difficult enterprise. The positive partial transposition (PPT) and moment-based criteria are two  key methods of entanglement detection. Among them, in an ideal set-up,  the PPT criterion can detect a greater volume of states but is more expensive, as it requires state tomography. We carry over the comparison of these two processes to a realistic situation with inaccurate detectors, in particular considering those that involve additional and lost counts. Considering Werner states, we find that the occurrence of additional counts does not result in false positives—declaring a separable state as entangled—for both  the methods. Additionally, we demonstrate that the PPT criterion detects a greater number of entangled Werner states compared to the moment-based method in this non-ideal scenario. Moreover, we find that detectors troubled by  lost counts may lead to false positives in both the methods. However, the moment-based criterion is error-free when the efficiency of the detectors is not significantly low; on the other hand, if the PPT criterion is used, a small deviation from the perfection of detectors can result in a wrong conclusion. The performance of the two criteria is further compared with both linear and nonlinear witness operators for Werner states. Nonlinear witness operators are found to be less accurate, whereas the linear witness achieves same accuracy as the PPT criterion followed by tomography. Finally, we consider Bell mixture states and two-qubit pure entangled states mixed with white noise. For both, we observe that no erroneous detection of entanglement occurs only in the presence of additional counts, whereas false detection arises in the presence of lost counts for both the PPT and $p_3$-PPT criteria.

\end{abstract}

\maketitle

\section{Introduction}
Entanglement is a type of quantum correlation  \cite{ent-review,G_hne_2009,ent2}, first observed by Einstein, Podolsky, and Rosen~\cite{EPR1}, which lacks any classical analogy. The discovery of entanglement opened a new path towards the success of quantum technology and communication~\cite{Teleportation, Densed, Keydistribution,Keyd2,Keyd1}. Its wide application includes protocols, such as quantum teleportation~\cite{tele}, and quantum devices, like quantum batteries~\cite{TT}. In these processes or systems, the presence of any finite non-zero amount of entanglement often provides a quantum advantage over the classical analog. To achieve this quantum advantage, it is crucial to at least correctly detect the presence of entanglement in a state, even if the exact amount of entanglement is not known. There exist various criteria for the detection of entanglement, such as the positive partial transposition (PPT) criterion~\cite{peres-sep,HorodeckiSep}, linear and nonlinear entanglement witnesses~\cite{witness2,G_hne_2009,Non-witness}, Bell inequality~\cite{Bell-inq} 
	
	The effectiveness of any particular criterion over the others for the detection of entanglement depends on the requirements. For example, the PPT criterion~\cite{necessery-sufficient} provides a necessary and sufficient condition for the detection of entanglement shared between two-qubit or qubit-qutrit systems. However, since partial transposition does not necessarily map positive operators to the same, it can not be experimentally implemented. Thus, the usual method to verify the criterion is to obtain the complete matrix form of the state through tomography~\cite{tomography1,tomo2,ar1,tomo4} and then analytically check the positiveness of the operator found by acting partial transposition on the tomographically obtained state. The process of tomography itself is expensive to implement in real experiments.
 Based on the same property, i.e., non-complete positiveness of partial transposition, another criterion for the detection of entanglement was introduced, which depends on the 2nd and 3rd moments of partial transposition of the shared state, the entanglement of which we want to detect.
 One of the major advantages of this new criterion lies in the fact that the second and third moments can be estimated in experiments without performing the costly method of complete state tomography \cite{p2_ppt3,p2_ppt2,P3ppt2,P3PPT-maint,mPT21}. 
	
	In experiments, the errors present in the apparatus can not always be completely ignored. These errors can have a significant impact on the result. Entanglement detection is also not an exception. Keeping in mind the inherent inaccuracies present in real-world experiments, we investigate the process of detecting entanglement. We consider an erroneous measurement process for entanglement detection. In particular, the motive is to observe the effect of the presence of detection errors in experiments. With this aim, we focus on distinct methods for the detection of entanglement, viz.,  PPT, moment-based, and linear and nonlinear entanglement witness based criteria. 

	
    The detection loophole is known  since 
    around 1970 from the works of Pearle, Santos and others~\cite{Pearle,Santos}. Its effects continue to pose challenges in the practical implementation of entanglement detection processes. 
Although detection loopholes have been studied in several entanglement detection schemes~\cite{Detloop11,Detloop12,Detloop13,LoopholeB,Bell3,Larsson_2014,Bell4,Bell5,Bell6,loophole1,loophole2,loophole3}, the presence of such loopholes in entanglement detection experiments using the moment-based and the PPT criteria has not been previously explored. For correct detection of entangled states using these criteria, the results available in earlier works are not adequate. We try to fill this gap in this work.
    
Within a measurement set-up, erroneous detectors can give rise to additional clicks in any particular measurement direction as well as can miss clicks in another or the same direction. This leads to the definition of two detector efficiencies: additional and lost event efficiencies. 
To examine the effect of the non-unit efficiencies corresponding to both types of error, we focus on the set of Werner states first and obtain the range of the parameter defining the Werner states for which a separable Werner state can appear to be entangled due to the inaccuracy of the detectors. Then the same study is also carried out for Bell mixture states and pure two-qubit entangled states mixed with white noise. 
	
	We first compare the regions of wrong detections, considering two detection conditions, which are the PPT and moment-based criteria, focusing on the Werner states. In this scenario, we find that, in the presence of lost events in the actual  detection (in a potential experiment) of entangled states using the moment-based criterion, there exists a threshold value of the lost event efficiency, above which, though a lower number of entangled states would be detected, no separable state would mistakenly appear to be entangled. But, in the case of PPT criterion, even a small deviation from the unity of the lost event efficiency will result in false detection. Even when both criteria fail to correctly detect the states, the volume of separable states being certified as entangled is smaller in the case of the moment-based criterion than the PPT criterion. Our findings highlight the fact that, besides the ease in experiments to estimate the moments, the $p_3$-PPT criterion is also beneficial for detecting entangled states in inaccurate experiments to avoid false positives in comparison to the PPT criterion.
 In the presence of additional counts, the effectiveness of the two criteria somewhat reverses. In such a scenario, the PPT criterion can correctly detect a finite volume of entangled Werner states for a wide range of values of efficiency and is free from erroneous detection, whereas the moment-based criterion is unable to certify any Werner state as entangled whenever the efficiency is less than 0.9. 
 
Furthermore, considering the set of Werner states, we compare the performance of PPT and $p_3$-PPT criteria with the efficacy of linear and nonlinear witness operators in the detection of entanglement. We first find that, in case of detection with witness operators also, erroneous results occur only for lost counts. We show, in the presence of such error, the performances of both the moment-based and PPT criteria are better than the performance of the nonlinear witness operator. Moreover, we find that the linear witness operator captures the same number of entangled states as the PPT criterion does for both types of error.
 
We followed the same analysis on mixture of Bell states and mixture of two-qubit pure entangled states with white noise, considering PPT and $p_3$-PPT criteria. We again find that there is no wrong detection in the presence of dark counts. Incorrect detection occurs only for lost counts for both the PPT criterion followed by state tomography and the $p_3$-PPT criterion.

The paper is structured as follows: In Sec.~\ref{sc2}, we describe the entanglement detection methods. In Sec. \ref{sc3}, we discuss the efficacy of entanglement detection methods in the presence of inaccuracies in the apparatus considering the PPT criterion, $p_3$-PPT criterion, and linear and nonlinear witnesses and make a comparative study between them. In Sec.~\ref{sc4}, we compare PPT and $p_3$-PPT criteria for Bell mixtures and non-maximally entangled states mixed with white noise. Finally, in Sec.~\ref{sc5}, we conclude.

\section{Detection of entanglement}\label{sc2}


There exist various methods for detection of entanglement. Using the PPT criterion, one can detect every entangled state that is not bound entangled. This makes the PPT criterion very powerful. However, to check the PPT criterion, the complete form of the quantum state needs to be determined, which can be done through state tomography. In contrast, the $p_3$-PPT criterion allows the detection of entanglement without requiring full knowledge of the quantum state. However, the $p_3$-PPT criterion is less powerful in the sense that it can, in general, detect fewer entangled states than the PPT criterion.
Linear and non-linear witnesses also do not need complete information about the state to detect the presence of entanglement in it. However, these criteria are useful when a state, in close proximity to the state that we want to detect, is known to be entangled. Otherwise, since there can be an infinite number of entanglement witnesses, it is difficult to choose the appropriate one that can detect the entanglement of the state in hand. 

Hence, depending on the available resources and knowledge about the state, an appropriate detection criterion needs to be chosen. There does not exist an ultimate detection criterion that overpowers all the others in every possible way. In this work, we compare the capability of the PPT criterion, the $p_3$-PPT criterion, and linear and nonlinear witnesses in entanglement detection, considering a realistic scenario.

Before going into the detail of the results, in the following subsections, we briefly recapitulate these detection processes.


{\subsection{PPT criterion}}

Partial transposition is a widely used method that provides necessary and sufficient criteria for detecting entanglement shared between qubit-qubit or qubit-qutrit systems.

	Let $\rho_{AB}$ be a bipartite quantum state shared between Alice and Bob. $\rho_{AB}$ acts on the Hilbert-space $\mathcal{H}_A\otimes\mathcal{H}_B$. Here the suffices $A$ and $B$ correspond to Alice and Bob's part respectively. 
	Let $\rho_{ab,cd}=\bra{a}\bra{b}\rho_{AB}\ket{c}\ket{d}$ be a particular element of the density matrix $\rho_{AB}$ where $\ket{a}$ ($\ket{b}$) and $\ket{c}$ ($\ket{d}$) are the elements of Alice's (Bob's) basis. The partial transposition of a state, $\rho_{AB}$, taken on Alice's subsystem, can be denoted by $\rho^{T_A}_{AB}$. 
	Under the action of the partial transposition on Alice's subsystem, any arbitrary element $\rho_{ab,cd}$ of the density matrix $\rho_{AB}$ transforms to $\rho_{c b,a d}$, i.e., the indices of Alice's subsystem gets swapped but Bob's indices remain unchanged. Transposition is a positive map, but it's not complete positive. Thus, though $\rho^T_{AB}$ (transposition on $\rho_{AB}$ over the composite basis of $\mathcal{H}_A\otimes\mathcal{H}_B$) is a positive operator, $\rho^{T_A}_{AB}$ may not be positive. This property gives birth to the entanglement detection criterion, which says if $\rho^{T_A}_{AB}$ is non-positive then $\rho_{AB}$ is surely entangled. The converse is also true in $2\otimes 2$ and $2\otimes 3$ systems but for higher dimensions, positivity of $\rho^{T_A}_{AB}$ does not confirm separability of $\rho_{AB}$. This is commonly known as the positive partial transpose criterion (PPT) or Peres Horodecki criterion~\cite{peres-sep,HorodeckiSep}.

The Werner state is given by $\rho_w=P\ket{\psi^{-}}\bra{\psi^{-}}+(1-P)\mathbbm{I}_4/4$, where $\ket{\psi^-}$ is the Bell singlet defined as $\ket{\psi^-}=(\ket{01}-\ket{10})/\sqrt{2}$ and $\mathbbm{I}_4$ is $4\times4$ identity matrix. The range of $P$ can be considered to be [0,1]. 

Therefore, the matrix form of $\rho_w$ in the computational basis is given by
	\begin{equation}
		\rho_w=\begin{bmatrix}
			(1-P)/4 & 0 & 0 & 0\\
			0 & (1+P)/4 & -P/2 & 0\\
			0 & -P/2 & (1+P)/4 & 0\\
			0 & 0 & 0 & (1-P)/4
		\end{bmatrix}. \nonumber
	\end{equation}
	If we take the partial transposition of the Werner state, $\rho_w$, with respect to the first party, say $A$, then we have
	\begin{equation}
		\rho_w^{T_A}=\begin{bmatrix}
			(1-P)/4 & 0 & 0 & -P/2\\
			0 & (1+P)/4 & 0 & 0\\
			0 & 0 & (1+P)/4 & 0\\
			-P/2 & 0 & 0 & (1-P)/4
		\end{bmatrix}. \nonumber
	\end{equation}.

The partially transposed density matrix is denoted by $\rho_w^{T_A}$. One can {easily check} that the four eigenvalues of $\rho_w^{T_A}$ are $(1+P)/4$, $(1+P)/4$, $(1+P)/4$, and $(1-3P)/4$. We see that three of the eigenvalues of $\rho^{T_A}_w$ are the same and positive for $0\leq P\leq 1$, but the eigenvalue $(1-3 P)/4 $ can be negative for a certain range of values of $P$. It can be checked that within the range $1/3<P\leq1$ the eigenvalue becomes negative. Hence for $1/3<P\leq1$ the Werner state contains entanglement. Since the PPT criterion provides a necessary and sufficient condition for the detection of entanglement for $2\otimes 2$ and $2\otimes 3$ states, the criterion confirms that the Werner state is separable for $P\leq 1/3$.

{\subsection{second and third moments}}
	
	In this section, we discuss about the PT-moments of a general bipartite state, which are another set of faithful quantities for detecting entanglement. In general, the $n$th PT-moment of a bipartite entangled state, $\rho_{AB}$, is defined as $p_n(\rho_{AB})=\text{Tr}\left[\left[(\rho_{AB})^{T_A}\right]^n\right]$, where $n$ is any positive integer defining the order of the PT-moment. $p_1(\rho_{AB})$ is always 1 for any state, $\rho_{AB}$, $p_2(\rho_{AB})$ quantifies the purity of $\rho_{AB}$, and $p_3(\rho_{AB})$ is the lowest order moment that carries information about the partial transposition taken over subsystem $A$.
	
	Let us mention an inequality using which we can detect the entanglement of any bipartite state. If the state $\rho_{AB}$ is separable, then $p_3(\rho_{AB})\geq p_2(\rho_{AB})^2$ and thus if $p_3(\rho_{AB})< p_2(\rho_{AB})^2$ we can surely certify the state, $\rho_{AB}$, as entangled \cite{P3PPT-maint,p2_ppt3,p2_ppt2,P3ppt2}. This condition is called the $p_3$-PPT criterion.\\

	For Werner state, $\rho_w$, $p_2(\rho_{w})$ and $p_3(\rho_w)$ are $(1+3P^2)/4$ and $(1+9P^2-6P^3)/4$, respectively. Using the $p_3$-PPT criterion a range of $P$ can be found beyond which
	$\rho_w$ becomes entangled and it is found to be $1/3<P\leq 1$. This range is the same with the range of $P$ that was obtained using PPT criterion.

\subsection{Linear entanglement witness}
Entanglement witness is another type of entanglement detection method that is introduced based on the Hahn-Banach theorem, where an operator is constructed to detect the entanglement \cite{lin_wit_1,lin_wit_2}. According to the theorem, corresponding to each entangled state, there always exists a linear operator that can detect the presence of the entanglement. Once this linear witness operator is found, the corresponding entangled state can be detected utilizing it. 

In general, a linear operator, $W$, is considered to be a witness operator if $\tr[\rho^{S}_{AB}W]\geq0$, for all separable states,  $\rho^S_{AB}$, that act on a fixed Hilbert space, and there exists at least one entangled state,  $\rho_{AB}$, in the same Hilbert space, for which $\tr[\rho_{AB}W]<0$. A well-known example of a linear witness operator that acts on $2\otimes 2$ systems is $W=\ket{\phi_+}\bra{\phi_+}^{T_B}$, where $\ket{\phi_+}=\frac{1}{\sqrt{2}}(\ket{00}+\ket{11})$. $W$ can detect the entanglement of states, $\rho^{\phi+}_{AB}$, for which the operator $\left(\rho^{\phi+}_{AB}\right)^{T_B}$ has a negative eigenvalue and the corresponding eigenvector is $\ket{\phi_+}$. Popular examples of such states are the Werner states, $\rho_w$.

\subsection{Nonlinear entanglement witness}

The process of entanglement detection using witness operators can be further improved by adding a nonlinear term with the linear witness operator, making the entire operator nonlinear~\cite{Non-lin_wit}. An example of such a nonlinear witness operator is
\begin{equation}
    F = \left\langle \, |\phi_+\rangle\langle\phi_+|^{T_B} \, \right\rangle
-
\frac{1}{s(\psi)}
\left\langle X^{T_B} \right\rangle
\left\langle \left(X^{T_B}\right)^\dagger \right\rangle.\label{eqq1}
\end{equation}
Here the first term is the expectation value of the linear witness operator discussed previously, which can detect Werner states. The operator, $X$, is defined as $X = |\phi_+\rangle\langle\psi|$, where $|\psi\rangle$ is an arbitrary but fixed state and $s(\ket{\psi})$ denotes the square of its largest Schmidt coefficient. The expectation values are taken over the state whose entanglement needs to be detected. The second term, which is subtracted from the linear term, is carefully chosen in such a way that it is always positive but still $ F\geq 0$ for all separable states. As a result, a negative value of $F$ indicates the presence of entanglement in the corresponding state on which $F$ is evaluated. Moreover, since $F\leq \langle W\rangle$ for all states, $F$ is stronger than the linear witness operator, $W$, in detecting entanglement. Hence, this nonlinear witness operator can also find all entangled Werner states.

 
\section{Detection loophole in Entanglement detection}\label{sc3}
Any theory is usually built depending on some idealistic assumptions. However, in real life, when we perform experiments, these assumptions may not remain valid. For example, there can always be some noise present in the apparatus that can deflect the experimental setup from it's theoretical structure. These unavoidable errors can affect the final conclusion of that experiment.

	Various kinds of loopholes can be present in entanglement detection experiments, for example the detection loophole~\cite{Detloop11,Detloop12,Detloop13,Bell3,Bell4,Bell5,Bell6,LoopholeB,loophole1,loophole2,loophole3}, locality loophole~\cite{loc-lop1}, coincidence loophole~\cite{coin-loop}, etc. Among them, we try to examine the effect of the presence of the detection loophole. In this context, we examine two particular detection methods one is based on the PPT criterion and the other one depends on the moment-based criterion. 

\subsection{PPT criterion}
 \label{SecIIIA}
Since partial transposition is not a physical operation, it can not be implemented experimentally. Thus the usual method of detection of entanglement using partial transposition criterion involves state tomography~\cite{tomography1,tomo2,ar1,tomo4}.
	\begin{figure}
		\centering

	\includegraphics[scale=0.36]{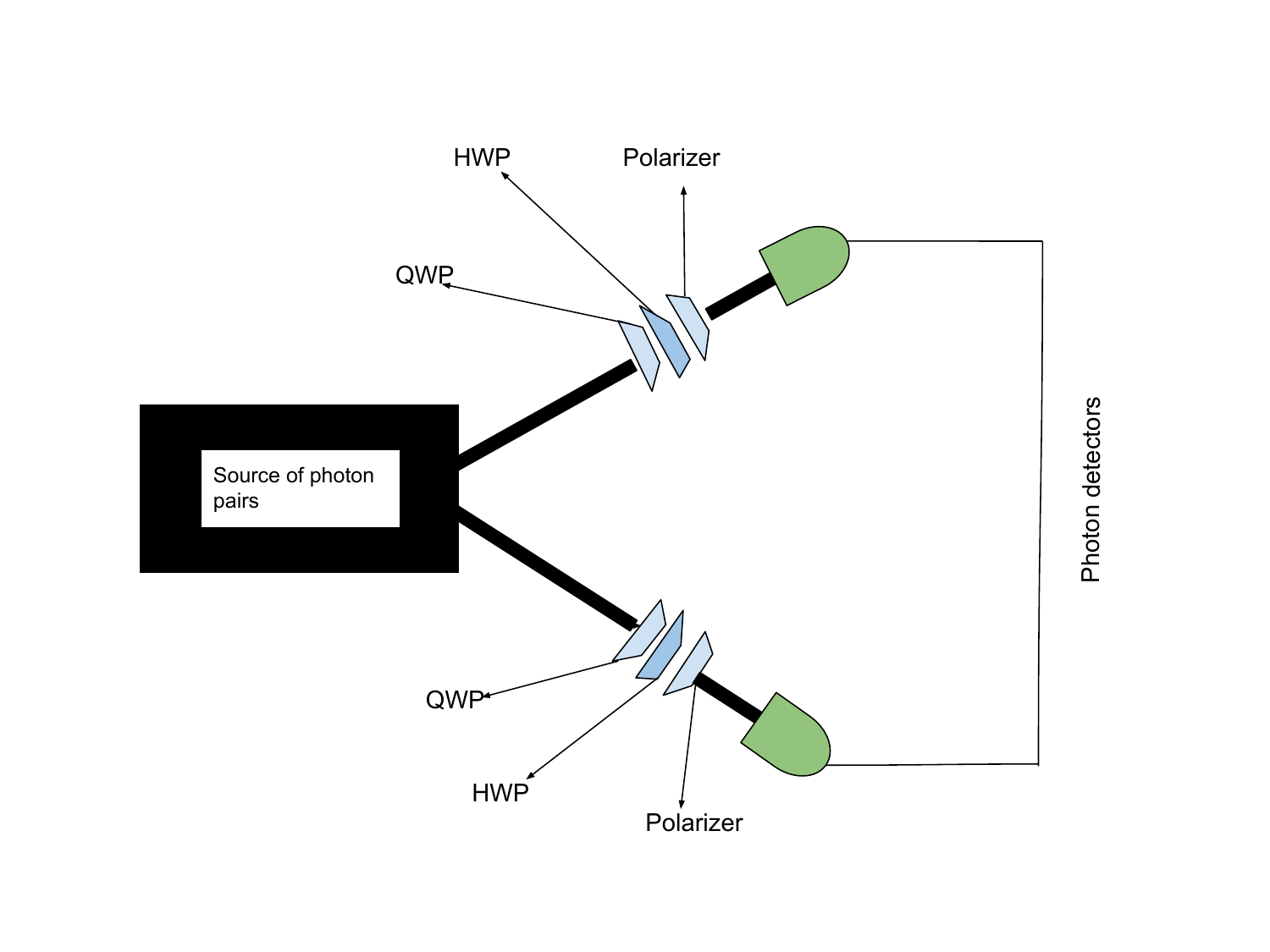}
		\caption{Schematic diagram of the experimental set-up, used for state tomography. Here QWP and HWP are the quarter and half wave plates, respectively, which are placed within the paths of the photons along with a polarizer. After the polarizer, the pair of photons passes through a pair of detectors where the polarization of the photons are measured~\cite{tomography1}.}
		\label{fig:my_label}
	\end{figure}
	
	The experimental arrangements required for quantum state tomography are depicted in Fig. 1. In general, a source of pairs of photons is used. Let $\ket{H}$ and $\ket{V}$ denote the two polarization states of each of the photons, i.e., horizontal and vertical, respectively.  Since $\ket{H}$ and $\ket{V}$ are two orthogonal states, $\{\ket{H},\ket{V}\}$ can form a basis of the Hilbert space describing the state of polarization of each of the photons. We can consider $\{\ket{H},\ket{V}\}$  as the computational basis. 
 From now on we will use this particular basis to represent any operator acting on the single-qubit Hilbert space. To express the polarization state of the two photons collectively we use composite basis $B=\{\ket{HH},\ket{HV},\ket{VH},\ket{VV}\}$. 
  The photons are maid to pass through a polarizer and quarter and half wave plates to project the light beams in a particular directions. Therefore the projected two-photon state would be $\ket{\psi_{\nu}}=\left[U_H\left(h_1^\nu\right)\otimes U_H\left(h_2^\nu\right)\right]\left[U_Q\left(q_1^\nu\right)\otimes U_Q\left(q_2^\nu\right)\right]\ket{VV}$, where $U_H(h)$ and $U_Q(q)$ denote the unitary operations representing the action of the half and quarter wave-plates,  respectively. The parameters, $h$ and $q$, are the angles made by the quarter and half wave plates to the vertical axis, respectively. The exact form of the unitaries are given by

        \begin{equation*}
		U_{Q}(q)=\begin{bmatrix}
			i-\cos(2q) & \sin(2q) \\
			\sin(2q) & i+\cos(2q) \\
		\end{bmatrix}
	\end{equation*}
		
	and

	\begin{equation*}
		U_{H}(h)=\begin{bmatrix}
			\cos(2h) & -\sin(2h) \\
			-\sin(2h) & -\cos(2h) \\
		\end{bmatrix}.
	\end{equation*}


 
 Let $\{\Gamma_{\mu}\}_{\mu}$ be a set of 16 linearly independent matrices, which satisfy the properties, 
    	{\begin{equation*}
		\Tr[\Gamma_{\mu},\Gamma_{\xi}]=\delta_{\mu\xi}~~~\text{and}~~~
		X=\sum_{\mu=1}^{16}\Gamma_{\mu}\Tr[\Gamma_{\mu} \cdot X].
	\end{equation*}

   Here $X$ is any arbitrary matrix. An example of such $\Gamma_{\mu}$ matrices are $\sigma_i\otimes\sigma_j$  for $i,j=0,1,2,3$. Here, $\sigma_1$,   $\sigma_2$, and $\sigma_3$ are Pauli matrices and  $\sigma_0$ is the 2$\times$2 identity operator. Any density matrix, $\rho$, can be expressed in terms of $\Gamma_{\mu}$ in the following way

  \begin{equation*}
		\rho=\sum_{\mu=1}^{16}r_{\mu}\Gamma_{\mu}.
	\end{equation*}
	From the trace relations of the Pauli matrices, we have 
	$r_{\mu}=\Tr[\Gamma_{\mu}  \rho].$
	Another 16$\times$16 dimensional matrix, $\mathcal{K}$, can be defined as
	\begin{equation}
		\mathcal{K}_{\nu,\mu}=\bra{\psi_{\nu}}\Gamma_{\mu}\ket{\psi_{\nu}}. \nonumber
	\end{equation}
	Therefore, number of counts in a particular direction $S^\nu$, can be expressed as
	\begin{equation}
		N_{\nu}=\mathcal{N}\sum_{\mu=1}^{16}\mathcal{K}_{\nu,\mu}r_{\mu}. \nonumber
	\end{equation}
	Hence the density matrix can be represented in terms of these experimentally accessible parameters, $N_\nu$, in the following way
	\begin{equation}
		\rho=(\mathcal{N}^{-1})\sum_{\nu=1}^{16}M_{\nu}N_{\nu}, \nonumber
	\end{equation}
        	where $M_{\nu}=\sum_{\mu=1}^{16}(\mathcal{K}^{-1})_{\nu,\mu}\Gamma_{\mu}$.
	The matrices, $M_\nu$, satisfy the completeness relation, $\sum_{\nu=1}^{16}M_{\nu}=\mathbbm{I}_4$ (For proof, see appendix of Ref. \cite{tomography1}).  To perform the tomography, one should select the sets, $S^\gamma$, in such a way that the inverse, $\mathcal{K}^{-1}$, exist.

Till now we have discussed the experiment by assuming an ideal setting. But in realistic situations, various type of noise, present in the apparatus, can significantly influence the results. Let us examine how the presence of imperfections in the detectors can affect the entanglement detection. Because of the imperfections, events can get lost or extra events can appear in the detector. In this inaccurate situation, the observed form of the density matrix will be 
	\begin{equation}
		\rho'=(\mathcal{N'}^{-1})\sum_{\nu=1}^{16}M_{\nu}N'_{\nu},\label{eq11}
	\end{equation}
	where $\mathcal{N'}$ is the total number of outcomes when measuring on a particular basis.

    Performance of an experimental set-up, in presence of errors, is best quantified by its efficiency. In the context of noisy scenario, two efficiencies can be defined, viz., the additional and lost event efficiencies, given by $\eta_+=\frac{\mathcal{N}}{\mathcal{N}+\varepsilon_+}$ and $\eta_-=\frac{\mathcal{N}-\varepsilon_-}{\mathcal{N}}$, respectively. Here $\varepsilon_+$, and $\varepsilon_-$ denote the total number of additional and lost events when the measurements are done on a particular basis. 
	We will consider two situations separately, in the first scenario we fix the additional event efficiency at $\eta_+=1$ and, in the second one, we consider $\eta_-=1$ with arbitrary $\eta_+$. 

	Let, because of the presence of imperfections in the detectors, the number of clicks in the direction $S^\nu$ is $N_{\nu}^{'}=N_{\nu}\pm\epsilon^{\pm}_{\nu},$
 where $N_\nu$ is the number of times it should have clicked in the ideal situation.
	Here, $+$ and $-$ sign correspond to additional and lost count type errors, respectively. To simplify the calculations and reduce notational complexity, we take $\epsilon^\pm_{\nu}=\epsilon_\pm$ for all $\nu$. In this noisy scenario, the observed form of the density matrix would be 
$     \rho'_{AB}=(\mathcal{N'}^{-1})\sum_{\nu=1}^{16}M_{\nu}N'_{\nu}=\left(\mathcal{N'}^{-1}\right)\left(\sum_{\nu=1}^{16}M_{\nu}N_{\nu}\pm\epsilon_{\pm}\mathbbm{I}_4\right)=\frac{\mathcal{N}}{\mathcal{N}'}\left(\rho_{AB}\pm\frac{\epsilon_\pm}{\mathcal{N}}\mathbbm{I}_4\right)$,
where $\mathcal{N'}$ is the total number of outcomes when measuring on a particular basis. 
 Here we have used the completeness relation of $M_\nu$. The entanglement of $\rho'_{AB}$ can be detected by observing whether any eigenvalue of the partially transposed density operator $\rho'^{T_A}_{AB}$ is negative or not. It is interesting to note that the positivity of $\rho'^{T_A}_{AB}$ is independent of not only the total number of outcomes, $\mathcal{N'}$, but also the chosen measurement basis, $M_\nu$, that is, in particular the angles, $\{h_1^\nu,h_2^{\nu},q_1^\nu,q_2^\nu\}$. 

	\textit{Case 1 : }
	To get a deeper understanding, we focus on a scenario where the actual shared state is a Werner state. In this case, we consider that counts can get lost in the measurement process but no additional counts will appear. 
 
 The experimentalists do not have any information about the shared state, other than it is a two-qubit state. They want to clarify if the shared state is entangled. The experimentalists consider a particular set of values for each of the angles, $h_1^\nu$, $h_2^\nu$, $q_1^\nu$, and $q_2^\nu$, such that the inverse of $\mathcal{K}$ exists. Because of the erroneous detectors, the matrix form they get is not exactly equal to $\rho_w$, but has a distinct structure,i.e, $\rho_w^-$, which is given by,

\begin{widetext}
		\begin{eqnarray}
			\rho^{-}_{w}=\frac{\mathcal{N}}{\mathcal{N'}}\begin{bmatrix}
				\frac{\mathcal{N}-P\mathcal{N}-4\epsilon_-}{4\mathcal{N}} & 0 & 0 & 0\\
				0 & \frac{\mathcal{N}+P\mathcal{N}-4\epsilon_-}{4\mathcal{N}} & -\frac{P}{2} & 0\\
				0 & -\frac{P}{2} & \frac{\mathcal{N}+P\mathcal{N}-4\epsilon_-}{4\mathcal{N}} & 0\\
				0 & 0 & 0 & \frac{\mathcal{N}-P\mathcal{N}-4\epsilon_-}{4\mathcal{N}}
			\end{bmatrix}.\nonumber
		\end{eqnarray}
	\end{widetext}
The smallest eigen-value of $(\rho_w^-)^{T_A}$ is $\frac{\mathcal{N}(1-3P)-4\epsilon_-}{4\mathcal{N'}}$. Therefore, if the experimentalists check the PPT criterion, they will certify all the states having $P$ within the range $1/3-4\epsilon_-/3\mathcal{N}<P\leq1$ as entangled. We see, for any given value of $\epsilon_-$ and $\mathcal{N}$, the states corresponding to the range $1/3-4\epsilon_-/3\mathcal{N}<P\leq\frac{1}{3}$ are separable, but will be falsely detected as entangled. Since in this case, the state under consideration is two-qubit, that is, the dimension of the Hilbert space is 4, $\varepsilon_-=4\epsilon_-$. Hence, the lost event efficiency is given by $\eta_{-}=(\mathcal{N}-4\epsilon_-)/\mathcal{N}.$ Therefore the range of $P$ for which the corresponding state would be detected as entangled can be expressed in terms of the lost event efficiency, $\eta_-$, and is given by, $\frac{\eta_-}{3}<P\leq1$. This implies $P_c=\eta_{-}/3$ is the cut-off value of the parameter, $P$, beyond which all states will appear as entangled in the erroneous method of entanglement detection.
Hence the states represented by $P\in[\frac{\eta_{-}}{3},\frac{1}{3}]$, though separable, would be certified as entangled, due to the presence of imperfections in the detector, which would lead to wrong detection of entanglement.

	\textit{Case 2 : }Let us now move to the next situation, where the lost event efficiency, $\eta_-$, is unity but the additional event efficiency, $\eta_+$, has an arbitrary value.
	In this scenario, because of over counts, $N_{\nu}^{'}=N_{\nu}+\epsilon_+.$
 Following the same method as discussed in the previous case, we get the range of $P$ for which the experimentalists will declare the Werner state as entangled, and it is given by $1/3+4\epsilon_+/3\mathcal{N}<P\leq1$.
	Since, by considering $\varepsilon_+=4\epsilon_+$, we have $\eta_+=\frac{\mathcal{N}}{\mathcal{N}+4\epsilon_+}$, the range of detected entangled states can be expressed in terms of the additional event efficiency, $\eta_+$, i.e., $\frac{1}{3\eta_+}<P\leq1$. Hence, in this case, the cut-off value of the parameter, $P$, above which all states will appear as entangled in the detection process is $P_c=\frac{1}{3\eta_+}$. We see, in this case, though a lesser number of entangled Werner states would be certified as entangled, no separable state would be falsely detected as entangled.

    
\subsection{Evaluation of $p_2$ and $p_3$ moments}

\begin{figure*}
			\includegraphics[width=0.44\textwidth]{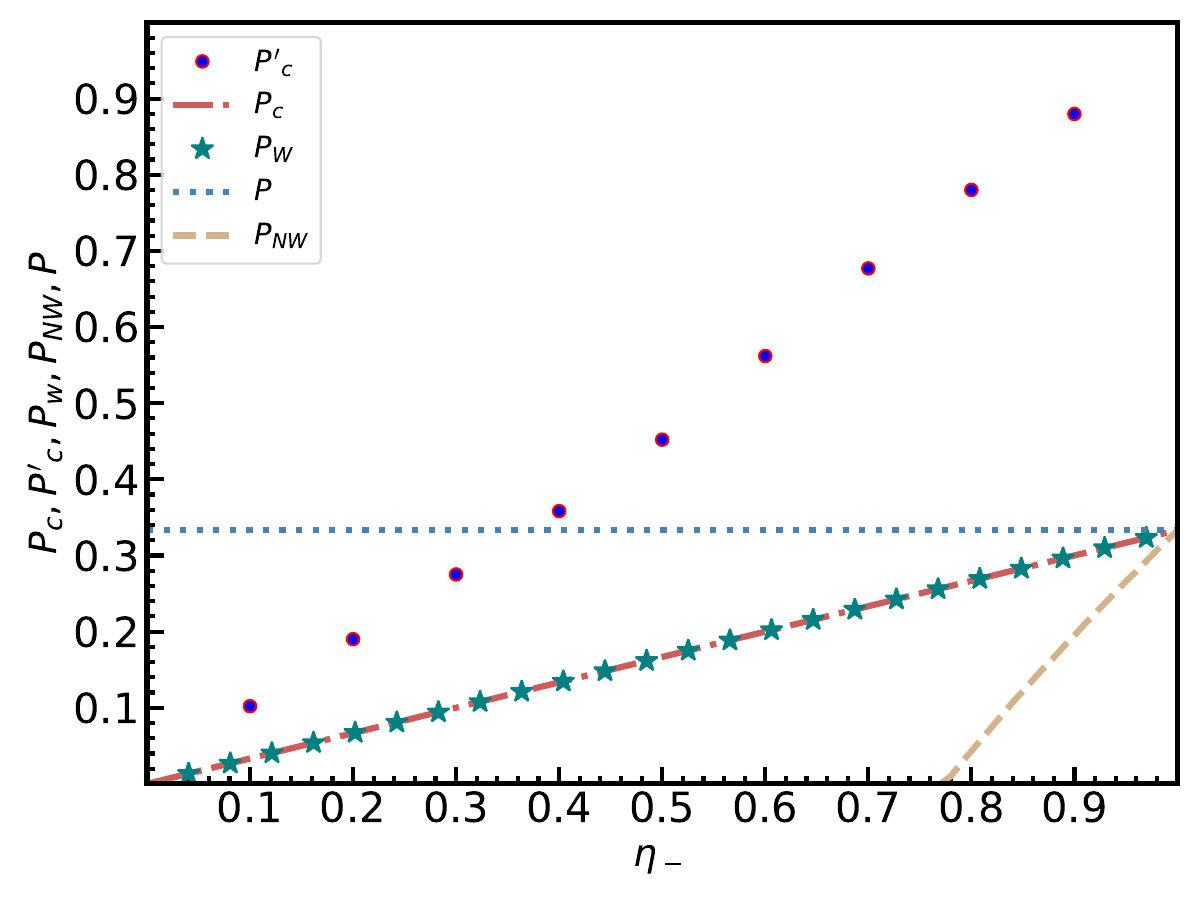}\hspace{15mm}
			\includegraphics[width=0.44\textwidth]{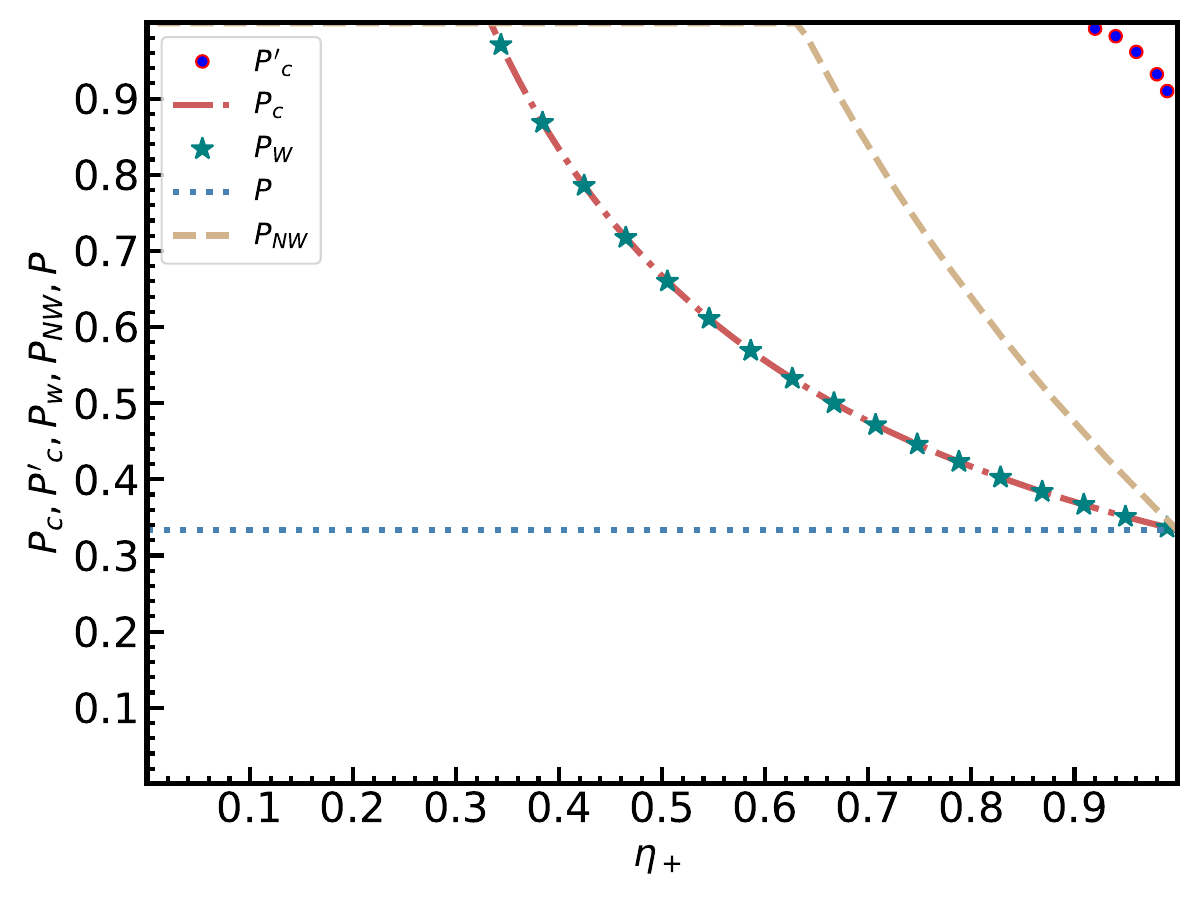}
		\caption{\textbf{Range of the parameter defining Werner states for which the state is certified as entangled.} We plot $P_c$, $P'_c$, $P_W$ and $P_{NW}$ using dash-dotted red, blue markers with red border,  green markers of star type, and dashed tan-colored lines, respectively, along the vertical axis in the presence of lost and additional counts, characterized by the lost event efficiency ($\eta_{-}$) and additional event efficiency ($\eta_{+}$) which are presented along the horizontal axis of left and right panels of the figure respectively. In both of the plots, the dashed sky blue line depicts the value of the parameter, $P$, (above which the state is entangled in a loophole-free scenario) of the Werner state above which the state truly becomes entangled. All the axes are dimensionless. }
		\label{fig:figures}	
	\end{figure*}
In this section, we first discuss about the detection of entangled states through evaluation of $p_2$ and $p_3$ moments of partial transposition \cite{P3PPT-maint}. Then we implement the same type of error {as implemented in the case of state tomography} and explore its effects. 
	
	Let us consider a $n$ qubit system. We make two partitions of this system, say $A$ and $B$. $|A|$ and $|B|$ are the number of qubits in the subsystems $A$ and $B$, respectively.
    
The usual method of experimental estimation of $p_2$ and $p_3$ moments of a state say $\rho_{AB}$, involves the operation of local random unitaries, $U_i$, on each of the qubit of $\rho_{AB}$. As a result, each of the qubits gets rotated arbitrarily with respect to each other. The composite unitary, acting on the complete $n$-qubit state, is given by $U=U_1\otimes U_2\otimes \cdot\cdot\cdot\otimes U_{n}$. After the operation of $U$, the state is projected on the computational basis. Let the outcome set of this projective measurement is $K=\{k_1,k_2,k_3, \cdot\cdot\cdot,k_{n}\}$. The combined operation of local unitaries and projective measurement can be done on, say, $M$ copies of $\rho_{AB}$. 
	The classical snapshots are defined as
    	\begin{equation}
		\hat{\rho}^{(r)}_{AB}=\ot_{i\in AB}\left[3(U_i^{(r)})^{\dagger}\ket{k^{(r)}_{i}}\bra{k^{(r)}_{i}}U^{(r)}_i-\mathbbm{I}_2\right], \nonumber
	\end{equation}
	where $U_i^{(r)}$ denotes the unitary operated on the $i$th qubit of the $r$th copy of the state, $\rho_{AB}$, $k_i^{(r)}$ is the outcome of the measurement on the same qubit of the same copy after the operation of the unitary, $U_i$, and $\mathbbm{I}_2$ is the identity matrix which operates on the individual qubits. The unbiased estimator of $l$th PT-moment, say $\hat{p}_l$, can be obtained in terms of all possible combinations of $l$ snapshots among the $M$ snapshots. The estimator is given by~\cite{Ustatistics} 
    \begin{eqnarray}
		\hat{p}_l=\frac{1}{l!}{\binom{M}{l}}^{-1}&\nonumber \\
		\sum_{r_1\neq r_2\neq.....\neq r_l} &\Tr[\Pi_A\Pi_B \hat{\rho}_{AB}^{(r_1)}\otimes\hat{\rho}_{AB}^{(r_2)}\otimes.....  \otimes\hat{\rho}_{AB}^{(r_l)}].\nonumber\\ \label{UEPN}
	\end{eqnarray} 
	
    Here $\Pi_A$ and $\Pi_B $ are the permutation operators, that is, 
    	\begin{eqnarray*}
		&&\Pi_A\ket{K_A^1,K_A^2,.....,K_A^l}=\ket{K_A^l,K_A^1,.....,K_A^{l-1}},\\
		&&\Pi_B\ket{K_B^1,.....,K_B^l}=\ket{K_B^2,K_B^3,.....,K_B^l,K_B^1,},
	\end{eqnarray*}
    
and $K_{A/B}^{(r)}=\{k_1^{(r)},k_2^{(r)}, \cdot\cdot\cdot,k_{|A|/|B|}^{(r)}\}$ denotes the set of measurement outcomes in A/B's side involved in the expression of $\hat{\rho}_{AB}^{(r)}$.

  Again we consider the Werner state, $\rho_w$.
   We operate random unitary on $M=2500$ copies of $\rho_w$. The projective measurement on each qubit can be done in $\{\ket{H}, \ket{V}\}$ basis. Since the Werner state is a two-qubit state, for each value of $r$ there can be four possible sets of measurement outcomes, $K^{r}$, i.e., $\{\ket{H},\ket{H}\}$, $\{\ket{H},\ket{V}\}$, $\{\ket{V},\ket{H}\}$, and $\{\ket{V},\ket{V}\}$. To implement the measurements theoretically, we randomly generated the set $K^r$ by following the probability distribution, $\bar{P}_\alpha=\bra{\alpha}U\rho_wU^\dagger\ket{\alpha}$, i.e., the probability of the state, $U\rho_w U^\dagger$, to get projected on $\ket{\alpha}$. Finally we calculate the unbiased estimator of $p_2$ and $p_3$ moments using Eq. \eqref{UEPN}. To find the exact values of $p_2$ and $p_3$, we repeat the process 100 times and take the average over the estimators. In the ideal case, the estimated values of $p^2_2$ and $p_3$ is found to be exactly equal to $(1+3P^2)^2/16$ and $(1+9P^2-6P^3)/4$, respectively, up to a numerical error which are of the order of $10^{-2}$. 
	To realize the effect of non-ideal detectors, we replace the probabilities, $\bar{P}_\alpha$, with $\bar{P}^-_{\alpha}=\frac{N_\alpha-\epsilon_{\alpha}^-}{\Tilde{N}}$ (considering lost events only) and $\bar{P}_{\alpha}^+=\frac{N_\alpha+\epsilon_{\alpha}^+}{\Tilde{N}}$ (considering only additional events). We again assume $\epsilon_{\alpha}^-=\epsilon_-$ and $\epsilon_{\alpha}^+=\epsilon_+$ for all $\alpha$.
	We can express $\bar{P}_{\nu}^-$ and $\bar{P}_{\nu}^+$ in terms of efficiency, $\eta_-$ and $\eta_+$, for lost and additional counts respectively, in the following way 
	\begin{equation*}
		\bar{P}_\alpha^-=\frac{(4\bar{P}_\alpha-(1-\eta_-))}{(4-4(1-\eta_-))}~~~\text{and}~~~\bar{P}_\alpha^+=\frac{(4\bar{P}_\alpha+(1/\eta_+-1))}{(4+4(1/\eta_+-1))}.
	\end{equation*}

 By varying the efficiencies, $\eta_{-}$ and $\eta_+$, we calculate the cut-off value of $P$, say $P'_c$, above which all the Werner states will be certified as entangled. To find this point, we calculate the values of $p_2^{2}$ and $p_3$ for different $P$ and particular values of efficiency, $\eta_\pm$, between the range $0$ to $1$. By plotting $p_3$ and $p_2^2$ with respect to $P$, for a fixed efficiency, $\eta_\pm$, we find the value of $P'_c$, that is, the value of $P$ at which $p_3$ and $p_2^2$ curves intersect. For the region $P>P_c'$, $p_3$ is less than $p_2^2$, indicating the state as entangled.

 In the left panel of Fig. \ref{fig:figures}, we plot $P'_c$ as a function of the lost event efficiency, $\eta_-$, taking $\eta_+=1$, using dots. To compare the method of detection of entangled states using $p_3$-PPT condition with PPT criterion, we plot $P_c$ in the same figure using a dash-dotted line. We also plot the actual value of $P$ above which the Werner state becomes entangled, i.e., $P=\frac{1}{3}$, using a blue dashed line. From the figure, it is apparent that there is no wrong detection in the moment-based method for efficiency above 0.4. But, in the case of state tomography, a minor deviation from the ideal scenario will result in wrong detection, that is certifying separable states as entangled. The higher slop of the curve joining the red pluses than the blue line confirms that even if the efficiency, $\eta_-$, is less than 0.4, the $p_3$-PPT condition approves fewer separable states as entangled than the combined method using state tomography and partial transposition. In the right panel of Fig. \ref{fig:figures}, we plot the same quantities but by varying $\eta_+$ and keeping $\eta_-$ fixed at one. It is visible from the figure that all the points  $P'_c$ and $P_c$ are above the dashed line. Therefore none of the two methods results in any wrong detection in the presence of additional counts. But the moment-based method can not detect any entangled state for $\eta_+\leq 0.9$, whereas the PPT-criterion can detect a finite amount of entangled states within the range $\frac{1}{3}<\eta_+\leq1$. Even when $\eta_+>0.9$ the moment-based criterion can detect a very small volume of entangled states compared to the PPT criterion.
	\subsection{Performance of linear witness}	
In this section, we compare the impact of noise on the detection processes using the linear and nonlinear entanglement witness methods, the moment-based criterion, and the PPT criterion checked after state tomography. 
\subsubsection{Linear witness operator}
Typically, a two-qubit linear witness operator can be decomposed in the Pauli basis as follows:
\begin{equation}
           W=\sum_{i,j=0}^3 C_{ij} \sigma_i\otimes\sigma_j=C_{00}\mathbb{I}_4+\sum_{k=1}^{15} C_kS_k.\label{eqq2}
\end{equation}
Here $C_{ij}$ and $C_k$ are real numbers, and $S_k$ denotes $\sigma_i\otimes \sigma_j$ for all $i=0,1,2,3$ and $j=0,1,2,3$, except $i=j=0$. To detect entanglement of a two-qubit state, $\rho_{AB}$, using the witness operator, $W$,  the state must be measured in the $\{S_k\}_{k}$ basis. The accurately measured value of $S_k$ in an ideal scenario is $\langle S_k\rangle =\frac{\sum_{\nu} N_{\nu}\lambda_{\nu}}{\mathcal{N}}$, where $N_{\nu}$ is the number of times the $\nu$th eigenvalue, $\lambda_{\nu}$, has occurred in the detector and $\mathcal{N}$ is the total number of outcomes. However, in practical scenarios, these measurements are prone to errors, making the true value of the operators significantly different from the measured value. We want to understand the effect of these imperfect measurements in the detection process.

As in the previous cases, we separately consider the emergence of dark and lost counts in the detectors. In the presence of dark and lost counts, the measured expectation values of $S_k$  take the forms 
\begin{eqnarray*}
    \langle S_k\rangle _+=\frac{\sum_{\nu}(N_\nu+\epsilon^+_{\nu})\lambda_{\nu}}{\mathcal{N}+ \varepsilon_+} \text{ and }
    \langle S_k\rangle _-=\frac{\sum_{\nu}({N}_{\nu}-\epsilon^-_{\nu})\lambda_{\nu}}{\mathcal{N}- \varepsilon_-},
\end{eqnarray*}
respectively. Here  $\epsilon^\pm_{\nu}$ represents the number of lost or additional counts of the $\nu$th eigenvalue, and $\varepsilon_{\pm}=\sum_\nu\epsilon^{\pm}_{\nu}$. These measured values, $\langle S_k\rangle +$ and $\langle Sk\rangle -$, can be expressed in terms of the efficiencies, $\eta+=\frac{\mathcal{N}}{\mathcal{N}+\varepsilon_+}$ and $\eta_-=\frac{\mathcal{N}-\varepsilon_-}{\mathcal{N}}$, which correspond to dark and lost counts, respectively, as $\langle S_k\rangle +={\eta+}\langle S_k\rangle$ and $\langle S_k\rangle -=\frac{1}{\eta-}\langle S_k\rangle$. Here the same assumptions on $\epsilon_{\nu}^{\pm}$ are considered as before. Hence, the measured expectation values of the witness operator, in the presence of dark and lost counts, are, respectively,
\begin{eqnarray*}
   \langle W\rangle _+&=&C_{00}\left(1-{\eta_+}\right)+{\eta_+}\langle W\rangle \text{ and} \\
    \langle W\rangle _-&=&C_{00}\left(1-\frac{1}{\eta_-}\right)+\frac{1}{\eta_-}\langle W\rangle.
\end{eqnarray*}
For the Werner state, $\rho_w$, and the witness operator, $W = \ket{\phi_+}\bra{\phi_+}^{T_B}$, the measured expectation values of $W$ reduce to 
\begin{equation*}
    \langle W\rangle _+ = \frac{1}{4}(1-3\eta_+P) \text{ and }\langle W\rangle _- = \frac{1}{4}\left(1 - \frac{3P}{\eta_-}\right). 
\end{equation*}
 From the condition for detection of entanglement, i.e., $\langle W\rangle _{\pm}<0$, the bound on the parameter, $P$, can be found over which the states would appear as entangled in the detection process. In this way, we find in the presence of dark counts there is no incorrect detection of entanglement; nevertheless, less entangled states would be detected. Specifically, if the efficiency drops below $\eta_+ = \frac{2}{3}$, no entangled state will be captured in the process.
 On the other hand, for lost counts we find even a slight deviation from the ideal scenario can lead to separable states' appearance as entangled, resulting in erroneous detection. It is also to be noted that the critical value of $P$($P_W$) over which the state is captured as entangled by the linear witness operator is the same as that obtained by the PPT criterion followed by state tomography for both lost and dark counts.  In Fig.~\ref{fig:figures}, the variation of the bound, $P_W$, on $P$, over which the states are detected as entangled, is shown as a function of $\eta_-$ (see panel (a)) and $\eta_+$ (check panel (b)) using green star type markers. It is clear from the plots that the performance of this linear witness operator in the presence of lost counts is worse in comparison to both PPT and $p_3$-PPT criteria, as it identifies more separable states as entangled.   

     \begin{figure*}
		\centering
\includegraphics[scale=0.25]{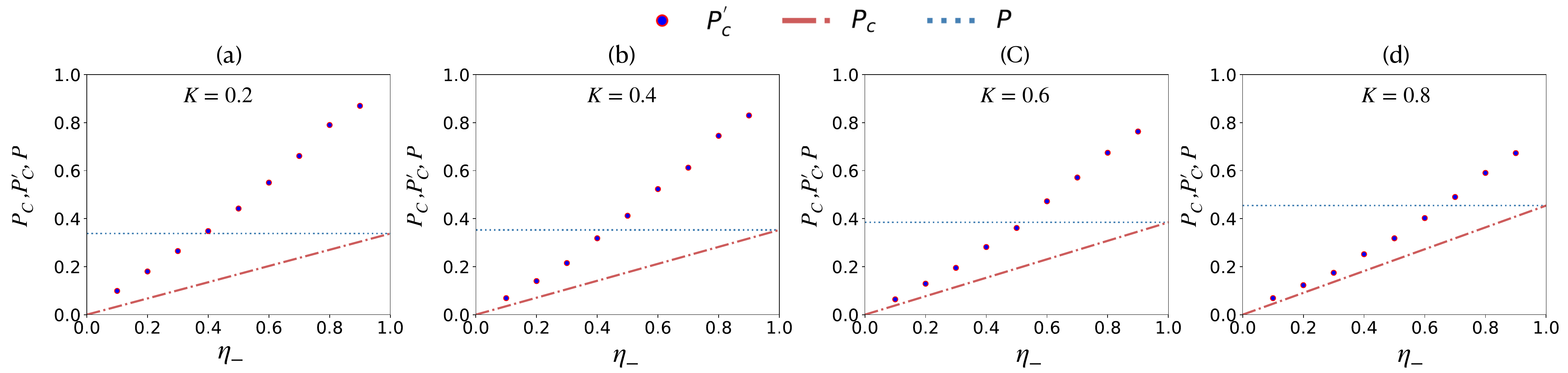}
\caption{\textbf{Detection of mixed states as entangled using $p_3$-PPT and PPT criteria, considering lost events in the measurement process.} Here, the horizontal axes represent the efficiency $(\eta_-)$ of the detectors, the non-unit value of which indicates lost events, while the vertical axes denote the bound on $P$, the parameter defining the state $\rho_K$, above which the state is identified as entangled. The entanglement detection is carried out using the PPT criterion, followed by state tomography, and the $p_3$-PPT method and the corresponding bounds on $P$ are denoted with $P_C$ and $P_C'$, respectively. The actual cut-off value of $P$ over which the state is truly entangled is denoted with $P$ itself. Panels (a), (b), (c), and (d) correspond to $K = 0.2$,  0.4, K = 0.6, and K = 0.8, respectively. 
}
\label{Poxpo}
\end{figure*}

\subsection{Detection with nonlinear witness}
Let us proceed to study the effect of the detection errors in entanglement detection using a nonlinear witness operator. In this regard, we focus on the nonlinear witness provided in Eq. \eqref{eqq1} considering $\ket{\psi}=\ket{00}$.}=
 
Since $X^{T_B}$, involved in the construction of nonlinear witnesses, is not in general Hermitian, for measurement purposes, it is decomposed in terms of two Hermitian operators, $H$ and $A$, as
$
X^{T_B} = H + iA.
$ Consequently, we get
$
\langle X^{T_B} \rangle \langle (X^{T_B})^\dagger \rangle
=
\langle H \rangle^2 + \langle A \rangle^2.
$
The form of the nonlinear witness in terms of these operators is $$F=\left\langle|\phi_+\rangle\langle\phi_+|^{T_B}\right\rangle-\frac{1}{s(\psi)}\left(\langle H \rangle^2 + \langle A \rangle^2 \right).$$ To detect the presence of entanglement in a state using this nonlinear function, both the measurable physical observables, $H$ and $A$, need to be measured, along with the linear operator $W= |\phi_+\rangle\langle\phi_+|^{T_B}$.

In a realistic scenario, the measured expectation values, $\langle W\rangle_m$, $\langle H\rangle_m$, and $\langle A\rangle_m$, would be in general different from their actual values, $\langle W\rangle$, $\langle H\rangle$, and $\langle A\rangle$, respectively. Following the same calculation as done for the linear witness operator, $W$, we find the relation of the measured values, $\langle H\rangle_m$ and $\langle A\rangle_m$, with the actual values, $\langle H\rangle$ and $\langle A\rangle$, to be
\begin{eqnarray*}
\langle H \rangle_m
&=&
C_{0H}\left(1 - \eta_+\right)
+
\eta_+\langle H \rangle\text{ and }\\
\langle A \rangle_m
&=&
C_{0A}\left(1 - {\eta_+}\right)
+
{\eta_+}\langle A \rangle,
\end{eqnarray*}
in the presence of dark counts. Similarly, for lost counts, the expressions can be evaluated to be
\begin{eqnarray*}
\langle H \rangle_m
&=&
C_{0H}\left(1 - \frac{1}{\eta_-}\right)
+
\frac{1}{\eta_-}\langle H \rangle\text{ and }\\
    \langle A \rangle_m
&=&
C_{0A}\left(1 - \frac{1}{\eta_-}\right)
+
\frac{1}{\eta_-}\langle A \rangle.
\end{eqnarray*}
Here $C_{0H}=\Tr[H]/4$ and $C_{0A}=\Tr[A]/4$ play the same role as $C_{00}$ in the decomposition of $W$ (see Eq. \eqref{eqq2}). Again, considering the measured value, $F_m$, of the nonlinear witness to be negative, we can find the bound, $P_{NW}$, on the parameter, $P$, above which the Werner state would be indicated as entangled in the detection process. 
In the left and right panels of Fig.~\ref{fig:figures}, we plot $P_{NW}$ using tan color for lost and dark counts, respectively. Similar to other considered detection methods, we find in the presence of dark counts there is no wrong entanglement detection. However, for lost counts making the efficiency belong within the range, $\eta_-\in[0.78,8.4]$, the nonlinear witness certifies more separable states as entangled than the linear one. For other values of $\eta_-$, both linear and nonlinear witness operators make the same amount of wrong detections, i.e., show the same separable states as entangled.
\section{Analysis considering other states}\label{sc4}

In this section, we discuss entanglement detection of other two-qubit states, viz., the Bell mixture state and the mixture of pure entangled states with white noise. 

\subsection{Two Bell mixture state}

We wish to compare the effectiveness of entanglement detection by both the $p_3$-PPT criterion and the PPT criterion followed by state tomography in the presence of dark and lost counts.

The Bell-mixture state is given by $\rho_B=P\ket{\psi_+}\bra{\psi_+}+(1-P)\ket{\psi_-}\bra{\psi_-}$, where $\ket{\psi_{\pm}}=\frac{1}{\sqrt{2}}(\ket{01}{\pm}\ket{10})$ and $P\in[0,1]$. The state is separable only at $P=0.5$; otherwise, it is entangled. If we account for the presence of dark and lost counts, characterized by the number of additional and missed events,  $\epsilon_{+}$ and $\epsilon_{-}$, respectively, the state constructed via tomography deviates from the ideal one. Let the measured erroneous state, in the presence of dark and lost counts, be $\rho^{+}_B$ and $\rho^{-}_B$, respectively. The set of eigenvalues of $\left(\rho^+_B\right)^{T_B}$ can easily be found to be $\{\frac{2\epsilon_++\mathcal{N}}{2\mathcal{N}'},\frac{2\epsilon_++\mathcal{N}}{2\mathcal{N}'},\frac{2\epsilon_++\mathcal{N}-2P\mathcal{N}}{2\mathcal{N}'},\frac{2\epsilon_+-\mathcal{N}+2P\mathcal{N}}{2\mathcal{N}'}\}$, where the notations have the same meaning as in Sec. \ref{SecIIIA}. The eigenvalues $\frac{2\epsilon_++\mathcal{N}-2P\mathcal{N}}{2\mathcal{N}'}$ and $\frac{2\epsilon_+-\mathcal{N}+2P\mathcal{N}}{2\mathcal{N}'}$ can be negative if $P>\frac{1}{2}+\frac{\epsilon_+}{\mathcal{N}}$ and $P<\frac{1}{2}-\frac{\epsilon_+}{\mathcal{N}}$. Therefore, the separable state would never be certified as entangled, whatever be the value of $\epsilon_+$. We find with the $p_3$-PPT criterion, also, no wrong detection occurs. 

On the other hand, if we consider the lost counts, where the amount of error in each measurement outcome is $\epsilon_-$, then in this case, the set of eigenvalues of the measured operator, $\left(\rho_B^-\right)^{T_B}$, found through tomography would be $\{\frac{2\epsilon_--\mathcal{N}}{2\mathcal{N}'},\frac{2\epsilon_--\mathcal{N}}{2\mathcal{N}'},\frac{-2\epsilon_--\mathcal{N}+2P\mathcal{N}}{2\mathcal{N}'},\frac{-2\epsilon_-+\mathcal{N}-2P\mathcal{N}}{2\mathcal{N}'}\}$. From here we can see that a state would be considered entangled if $P$ belongs in the range $\frac{1}{2}-\frac{\epsilon_-}{\mathcal{N}}<P<\frac{1}{2}+\frac{\epsilon_-}{\mathcal{N}}$. Hence, even a single lost count will result in wrong detection of entanglement because the state is separable at $P=0.5$. The situation remains the same even if the $p_3$-PPT criterion is used for detection.

\subsection{Pure two-qubit entangled states mixed with white noise}
In this section, we focus on comparing the effectiveness of the PPT criterion implemented via state tomography and the $p_3$–PPT criterion in detecting entanglement of entangled states mixed with white noise, under realistic scenarios. The state under consideration is given by
\begin{equation}
    \rho_K=P\ket{\psi_K}\bra{\psi_K}+(1-P)\mathbbm{I}_4/4,
\end{equation}
where $\ket{\psi_K}=\sqrt{\frac{1+K}{2}}\ket{00}+\sqrt{\frac{1-K}{2}}\ket{11}$ and $K\in[-1,1]$. For $K=-1$ and $K=1$, the state, $\ket{\psi_K}$, is product, and otherwise is entangled. The state, $\rho_K$, is entangled for $P>\frac{1}{1 + 2\sqrt{1 - k^2}}$.  

We consider the scenario of detecting if the state, $\rho_K$, is entangled using the mentioned criteria. In Fig.~\ref{Poxpo}, we plot the bound on $P$ above which the state appears as entangled in the detection process, with respect to $\eta_-$, considering $p_3$-PPT criterion and state tomography with PPT criterion. The notations $P_C$, $P'_C$, and $P$, used in the plots, have the same meaning as in the plots of Fig.~\ref{fig:figures}.
From the figure, it is clear that there exists a threshold value of efficiency, below which the wrong detection of entanglement occurs with $p_3$-PPT criterion. This critical value depends on $K$ and is 0.40, 0.43, 0.52, and 0.63 for $K=0.2$, 0.4, 0.6, and 0.8, respectively. In case of PPT criterion, just a slight deviation from the ideal situation results in imperfect detection of separable states as entangled for lost count type error. For dark count type loophole, there is no wrong detection of entanglement. 

\textbf{Remark.} The entire analysis in this work is based on the consideration that the efficiency of the detectors used in the laboratory is known to the experimentalists. This is a standard requirement that can be met by performing the detection experiment on known states and checking how much the measured data differs from the actual data of the input state. However, even if the exact detection efficiency is not accurately known, our analysis can still be utilized, as long as a lower bound on the efficiency is available, by considering the worst-case scenario and assuming the minimum possible efficiency.

\section{Conclusion}\label{sc5}
Theories always begin with a set of assumptions that provide an idealistic concept about the world. But in reality, events do not always fall into the assumptions, which may lead to violations. The entanglement detection process is not an exception. Though there are multiple methods for the detection of entanglement, most of them are constructed considering an ideal situation where the apparatus involved in the experiment is taken to be perfectly efficient and isolated or non-interacting with the rest of the world. In physical scenarios, these assumptions are hard to maintain, and as a result of the inaccuracies that may affect the experiment, the observer may face false detection of entangled states. We know entangled states, gifted with the most unique quantum resource, provide an advantage in numerous technological and communicational tasks. Hence, the identification of separable states as entangled can result in serious damage to the effectiveness of these tasks.

Two of the most useful methods of entanglement detection are state tomography, followed by the application of the PPT criterion, and the usage of the $p_3$-PPT criterion. It is known that the implementation of the $p_3$-PPT criterion in the laboratory is easier because it does not require the entire state tomography as is needed for checking the PPT criterion. On the other hand, the latter can detect a larger volume of entangled states than the former. In this work, we want to compare these two methods considering a non-ideal experimental set-up.


  In ideal situations, both of these criteria are known to be capable of detecting all entangled Werner states. To allow inaccuracies we have considered the presence of additional counts and the absence of true counts in the measurements involved in the detection process. We realize that, for both criteria, allowing additional clicks in the measurements, involved in the entanglement-detection, experiment results in no incorrect detection. But in that scenario, less entangled states would be detected using the moment-based criterion compared to the PPT criterion. When we only considered lost counts in the measurement, we encountered wrong detection; that is, in such a situation, the experiment may certify separable states as entangled. However, the range of the parameter representing the wrong detection for a particular efficiency is smaller using the $p_3$-PPT criterion than the PPT criterion. Moreover, after a threshold value of the efficiency, the moment-based criterion becomes detection-loophole-free, whereas the PPT criterion can still erroneously detect states even when the efficiency is slightly lower than the perfect efficiency.

Thus, we can conclude that detecting entanglement through moments of partial transposition is more reliable than doing full-state tomography and checking the positivity of the partially transposed state, when the motive is the correct - unambiguous - detection of entanglement in quantum states. On the other hand, if the aim is to detect entanglement of a larger volume of states, the PPT criterion is more beneficial, but it comes with the drawback of claiming false detection. In addition to the simplicity of calculating moments in experiments, our results highlight that the $p_3$-PPT criterion provides a crucial advantage in entangled state detection in noisy scenarios. Compared to the entanglement detection method based on the PPT condition, the moment-based criterion results in reduced false positives.

Moreover, we compare the performance of PPT and $p_3$-PPT criteria with linear and nonlinear witness operators in entangled-state detection. We find the nonlinear witness operator makes more wrong detections than the PPT as well as the $p_3$-PPT conditions. It is also to be noted that the same states are detected wrongly in both the linear witness method and the PPT criterion followed by state tomography.

Even by considering both Bell mixture states and entangled states mixed with white noise, we observe that no erroneous detection occurs in the presence of dark-count-type errors. For both classes of states, incorrect detection is observed only in the presence of lost count type errors for both the PPT criterion followed by state tomography and the $p_3$-PPT criterion.

By exploring the detection of these multiple sets of states, we see the qualitative nature of the detection process in the presence of errors does not significantly depend on the state being examined or any particular property of that state. Therefore, we expect the overall understanding about the detection methods gained through this work will remain the same even if other states are examined.

\section{DATA AVAILABILITY}
 The data that support the findings of this article are openly
available~\cite{chaki_2026_20040539}.

\acknowledgements
Kornikar Sen acknowledges support from the project MadQ-CM (Madrid Quantum de la Comunidad de Madrid) funded by the European Union (NextGenerationEU, PRTRC17.I1) and by the Comunidad de Madrid (Programa de Acciones Complementarias). The research of
PC was supported by the INFOSYS scholarship.

\bibliography{P3PPT}

@article{ent-review,
  title = {Quantum entanglement},
  author = {Horodecki, R. and Horodecki, P. and Horodecki, M. and Horodecki, K.},
  journal = {Rev. Mod. Phys.},
  volume = {81},
  issue = {2},
  pages = {865},
  numpages = {0},
  year = {2009},
  month = {Jun},
  publisher = {American Physical Society},
  url = {https://link.aps.org/doi/10.1103/RevModPhys.81.865}
}

@article{peres-sep,
  title = {Separability Criterion for Density Matrices},
  author = {Peres, A.},
  journal = {Phys. Rev. Lett.},
  volume = {77},
  issue = {8},
  pages = {1413},
  numpages = {0},
  year = {1996},
  month = {Aug},
  publisher = {American Physical Society},
  url = {https://link.aps.org/doi/10.1103/PhysRevLett.77.1413}
}

@article{HorodeckiSep,
title = {Separability of mixed states: necessary and sufficient conditions},
journal = {Phys. Lett. A},
volume = {223},
number = {1},
pages = {1},
year = {1996},
issn = {0375-9601},
url = {https://www.sciencedirect.com/science/article/pii/S0375960196007062},
author = { Horodecki, M. and  Horodecki, P. and  Horodecki, R.}
}

@article{Teleportation,
  title = {Teleporting an unknown quantum state via dual classical and Einstein-Podolsky-Rosen channels},
  author = {Bennett, C. H. and Brassard, G. and Cr\'epeau, C. and Jozsa, Richard, H. and Peres, A. and Wootters, W. K.},
  journal = {Phys. Rev. Lett.},
  volume = {70},
  issue = {13},
  pages = {1895},
  numpages = {0},
  year = {1993},
  month = {Mar},
  publisher = {American Physical Society},
   url = {https://link.aps.org/doi/10.1103/PhysRevLett.70.1895}
}

@article{Densed,
  title = {Communication via one- and two-particle operators on Einstein-Podolsky-Rosen states},
  author = {Bennett, C. H. and Wiesner, S. J.},
  journal = {Phys. Rev. Lett.},
  volume = {69},
  issue = {20},
  pages = {2881},
  numpages = {0},
  year = {1992},
  month = {Nov},
  publisher = {American Physical Society},
  url = {https://link.aps.org/doi/10.1103/PhysRevLett.69.2881}
}

@article{Keydistribution,
  title = {Quantum cryptography based on Bell's theorem},
  author = {Ekert, A. K.},
  journal = {Phys. Rev. Lett.},
  volume = {67},
  issue = {6},
  pages = {661},
  numpages = {0},
  year = {1991},
  month = {Aug},
  publisher = {American Physical Society},
  url = {https://link.aps.org/doi/10.1103/PhysRevLett.67.661}
}

@book{Bell-inq,
  title={Speakable and unspeakable in quantum mechanics: Collected papers on quantum philosophy},
  author={Bell, J. S.},
  year={2004},
  publisher={Cambridge university press}
}

@article{P3PPT-maint,
  title = {Mixed-State Entanglement from Local Randomized Measurements},
  author = {Elben, A. and Kueng, R. and Huang, H. Y. R. and van Bijnen, R. and Kokail, C. and Dalmonte, M. and Calabrese, P. and Kraus, B. and Preskill, J. and Zoller, P. and Vermersch, B.},
  journal = {Phys. Rev. Lett.},
  volume = {125},
  issue = {20},
  pages = {200501},
  numpages = {6},
  year = {2020},
  month = {Nov},
  publisher = {American Physical Society},
  url = {https://link.aps.org/doi/10.1103/PhysRevLett.125.200501}
}

@article{Non-witness,
  title = {Nonlinear Entanglement Witnesses},
  author = {G\"uhne, O. and L\"utkenhaus, N.},
  journal = {Phys. Rev. Lett.},
  volume = {96},
  issue = {17},
  pages = {170502},
  numpages = {4},
  year = {2006},
  month = {May},
  publisher = {American Physical Society},
  url = {https://link.aps.org/doi/10.1103/PhysRevLett.96.170502}
}

@article{tomography1,
  title = {Measurement of qubits},
  author = {James, D. F. V. and Kwiat, P. G. and Munro, W. J. and White, A. G.},
  journal = {Phys. Rev. A},
  volume = {64},
  issue = {5},
  pages = {052312},
  numpages = {15},
  year = {2001},
  month = {Oct},
  publisher = {American Physical Society},
  url = {https://link.aps.org/doi/10.1103/PhysRevA.64.052312}
}

@article{loophole1,
  title = {Closing the detection loophole in nonlinear entanglement witnesses},
  author = {Sen, K. and Das, S. and Sen, U.},
  journal = {Phys. Rev. A},
  volume = {100},
  issue = {6},
  pages = {062333},
  numpages = {6},
  year = {2019},
  month = {Dec},
  publisher = {American Physical Society},
 
  url = {https://link.aps.org/doi/10.1103/PhysRevA.100.062333}
}

@article{loophole2,
  title = {Detection loophole in measurement-device-independent entanglement witnesses},
  author = {Sen, K. and Srivastava, C. and Mal, S. and Sen, A. and Sen, U.},
  journal = {Phys. Rev. A},
  volume = {103},
  issue = {3},
  pages = {032415},
  numpages = {7},
  year = {2021},
  month = {Mar},
  publisher = {American Physical Society},
  url = {https://link.aps.org/doi/10.1103/PhysRevA.103.032415}
}

@article{loophole3,
  title = {Noisy quantum input loophole in measurement-device-independent entanglement witnesses},
  author = {Sen, K. and Srivastava, C. and Mal, S. and Sen, A. and Sen, U.},
  journal = {Phys. Rev. A},
  volume = {104},
  issue = {1},
  pages = {012429},
  numpages = {8},
  year = {2021},
  month = {Jul},
  publisher = {American Physical Society},
  url = {https://link.aps.org/doi/10.1103/PhysRevA.104.012429}
}

@incollection{Ustatistics,
  title={A class of statistics with asymptotically normal distribution},
  author={Hoeffding, W.},
  booktitle={Breakthroughs in statistics},
  pages={308},
  year={1992},
  publisher={Springer}
}

@article{loc-lop1,
    title = {Random 'choices' and the locality loophole},
     author = {Pironio, S.},
     journal = {arXiv:1510.00248},
      year = {2015},
   url = {https://arxiv.org/abs/1510.00248},
  }

@article{coin-loop,
	
  
	url = {https://doi.org/10.1209%2Fepl%2Fi2004-10124-7},
  
	year = 2004,
	month = {sep},
  
	publisher = {{IOP} Publishing},
  
	volume = {67},
  
	number = {5},
  
	pages = {707},
  
	author = {J.-{\AA} Larsson and R. D Gill},
  
	title = {Bell{\textquotesingle}s inequality and the coincidence-time loophole},
  
	journal = {Euro. phys. Lett. ({EPL})}
}

@article{ent2,
      title = {The separability versus entanglement problem},
      author = {Das, S. and Chanda, T. and Lewenstein, M. and Sanpera, A. and De, A Sen. and Sen, U.},
      journal = {arXiv:1701.02187},
      year = {2017},
      url = {https://arxiv.org/abs/1701.02187},
}

@article{P3ppt2,
  title = {Machine-Learning-Assisted Many-Body Entanglement Measurement},
  author = {Gray, J. and Banchi, L. and Bayat, A. and Bose, S.},
  journal = {Phys. Rev. Lett.},
  volume = {121},
  issue = {15},
  pages = {150503},
  numpages = {6},
  year = {2018},
  month = {Oct},
  publisher = {American Physical Society},
 
  url = {https://link.aps.org/doi/10.1103/PhysRevLett.121.150503}
}

@article{tomo2,
  title = {Experimental Single-Setting Quantum State Tomography},
  author = {Stricker, R. and Meth, M. and Postler, L. and Edmunds, C. and Ferrie, C. and Blatt, R. and Schindler, P. and Monz, T. and Kueng, R. and Ringbauer, M.},
  journal = {PRX Quantum},
  volume = {3},
  issue = {4},
  pages = {040310},
  numpages = {34},
  year = {2022},
  month = {Oct},
  publisher = {American Physical Society},
  
  url = {https://link.aps.org/doi/10.1103/PRXQuantum.3.040310}
}

@article{ar1,
 title = {Quantum tomography using state-preparation unitaries},
 author = {van Apeldoorn, J. and Cornelissen, A. and Gilyén, A. and Nannicini, G.},
  journal = {arXiv:2207.08800},
  
  year = {2022},
  
  url = {https://arxiv.org/abs/2207.08800},
  
  

}

@article{tomo4,
 title = {Quantum State Tomography Inspired by Language Modeling},
 author = {Zhong, L. and Guo, C. and Wang, X.},
  journal = {arXiv:2212.04940},
  
  year = {2022},
  
  url = {https://arxiv.org/abs/2212.04940},
}

@article{necessery-sufficient,
	
  
	url = {https://doi.org/10.1016%2Fs0375-9601%2896%2900706-2},
  
	year = 1996,
	month = {nov},
  
	publisher = {Elsevier {BV}
},
  
	volume = {223},
  
	number = {1-2},
  
	pages = {1},
  
	author = {Horodecki, M. and  Horodecki, P. and Horodecki, R.},
  
	title = {Separability of mixed states: necessary and sufficient conditions},
  
	journal = {Phys. Lett. A.}
}

@article{p2_ppt3,
  title = {Quantifying entanglement of a two-qubit system via measurable and invariant moments of its partially transposed density matrix},
  author = {Bartkiewicz, K. and Beran, J. and Lemr, K. and Norek, M. and Miranowicz, A.},
  journal = {Phys. Rev. A},
  volume = {91},
  issue = {2},
  pages = {022323},
  numpages = {10},
  year = {2015},
  month = {Feb},
  publisher = {American Physical Society},
  
  url = {https://link.aps.org/doi/10.1103/PhysRevA.91.022323}
}

@article{EPR1,
  title = {Can Quantum-Mechanical Description of Physical Reality Be Considered Complete?},
  author = {Einstein, A. and Podolsky, B. and Rosen, N.},
  journal = {Phys. Rev.},
  volume = {47},
  issue = {10},
  pages = {777},
  numpages = {0},
  year = {1935},
  month = {May},
  publisher = {American Physical Society},
 
  url = {https://link.aps.org/doi/10.1103/PhysRev.47.777}
}

@article{G_hne_2009,
	
  
	url = {https://doi.org/10.1016%2Fj.physrep.2009.02.004},
  
	year = 2009,
	month = {apr},
  
	publisher = {Elsevier {BV}
},
  
	volume = {474},
  
	number = {1-6},
  
	pages = {1},
  
	author = {G\"uhne, O. and G{\'{e}}za, T.},
  
	title = {Entanglement detection},
  
	journal = {Phys. Rep.}
}

@article{Bell3,
  title = {Using macroscopic entanglement to close the detection loophole in Bell-inequality tests},
  author = {Lim, Y. and Paternostro, M. and Kang, M. and Lee, J. and Jeong, H.},
  journal = {Phys. Rev. A},
  volume = {85},
  issue = {6},
  pages = {062112},
  numpages = {5},
  year = {2012},
  month = {Jun},
  publisher = {American Physical Society},

  url = {https://link.aps.org/doi/10.1103/PhysRevA.85.062112}
}

@article{Bell4,
  title = {Closing the detection loophole in tripartite Bell tests using the $W$ state},
  author = {P\'al, K. and V\'ertesi, T.},
  journal = {Phys. Rev. A},
  volume = {92},
  issue = {2},
  pages = {022103},
  numpages = {9},
  year = {2015},
  month = {Aug},
  publisher = {American Physical Society},

  url = {https://link.aps.org/doi/10.1103/PhysRevA.92.022103}
}

@article{Bell5,
  title = {Device-Independent Bounds on Detection Efficiency},
  author = {Szangolies, J. and Kampermann, H. and Bru\ss{}, D.},
  journal = {Phys. Rev. Lett.},
  volume = {118},
  issue = {26},
  pages = {260401},
  numpages = {5},
  year = {2017},
  month = {Jun},
  publisher = {American Physical Society},

  url = {https://link.aps.org/doi/10.1103/PhysRevLett.118.260401}
}

@article{Bell6,
  title = {Complete list of tight Bell inequalities for two parties with four binary settings},
  author = {Cruzeiro, E. Z. and Gisin, N.},
  journal = {Phys. Rev. A},
  volume = {99},
  issue = {2},
  pages = {022104},
  numpages = {7},
  year = {2019},
  month = {Feb},
  publisher = {American Physical Society},
 
  url = {https://link.aps.org/doi/10.1103/PhysRevA.99.022104}
}

@article{Keyd1,
  title = {Simple Proof of Security of the BB84 Quantum Key Distribution Protocol},
  author = {Shor, P. W. and Preskill, J.},
  journal = {Phys. Rev. Lett.},
  volume = {85},
  issue = {2},
  pages = {441},
  numpages = {0},
  year = {2000},
  month = {Jul},
  publisher = {American Physical Society},

  url = {https://link.aps.org/doi/10.1103/PhysRevLett.85.441}
}

@article{Keyd2,
  title = {Optimal Eavesdropping in Quantum Cryptography with Six States},
  author = {Bru\ss{}, D.},
  journal = {Phys. Rev. Lett.},
  volume = {81},
  issue = {14},
  pages = {3018},
  numpages = {0},
  year = {1998},
  month = {Oct},
  publisher = {American Physical Society},

  url = {https://link.aps.org/doi/10.1103/PhysRevLett.81.3018}
}

@article{mPT21,
	
	url = {https://doi.org/10.1038%2Fs41534-021-00487-y},
  
	year = 2021,
	month = {oct},
  
	publisher = {Springer Science and Business Media {LLC}
},
  
	volume = {7},
  
	number = {1},
  
	author = {Neven, A. and Carrasco, J. and Vitale, V. and Kokail, C. and Elben, A. and Dalmonte, M. and Calabrese, P. and Zoller, P. and Vermersch, B. and  Kueng, R. and Kraus, B.},
  
	title = {Symmetry-resolved entanglement detection using partial transpose moments},
  
	journal = {npj Quantum Information}
}

@article{p2_ppt2,
    title = {Estimating the Entanglement Negativity from low-order moments of the partially transposed density matrix},
    author = {Carteret, H. A.},
    journal = {arXiv:1605.08751},
    year={2016},

    url = {https://arxiv.org/abs/1605.08751}
}

@article{Detloop11,
  title = {Hidden-Variable Example Based upon Data Rejection},
  author = {Pearle, P.},
  journal = {Phys. Rev. D},
  volume = {2},
  issue = {8},
  pages = {1418},
  numpages = {0},
  year = {1970},
  month = {Oct},
  publisher = {American Physical Society},
 
  url = {https://link.aps.org/doi/10.1103/PhysRevD.2.1418}
}

@article{Detloop12,
  title = {Experimental consequences of objective local theories},
  author = {Clauser, J. F. and Horne, M.},
  journal = {Phys. Rev. D},
  volume = {10},
  issue = {2},
  pages = {526},
  numpages = {0},
  year = {1974},
  month = {Jul},
  publisher = {American Physical Society},

  url = {https://link.aps.org/doi/10.1103/PhysRevD.10.526}
}

@article{Detloop13,
  title = {Critical analysis of the empirical tests of local hidden-variable theories},
  author = {Santos, E.},
  journal = {Phys. Rev. A},
  volume = {46},
  issue = {7},
  pages = {3646},
  numpages = {0},
  year = {1992},
  month = {Oct},
  publisher = {American Physical Society},

  url = {https://link.aps.org/doi/10.1103/PhysRevA.46.3646}
}

@article{witness2,
author = {    Bruß, D.  and  Cirac, J. I.    and     Horodecki, P.  and     Hulpke, F. and    Kraus, B. and     Lewenstein, M.  and     Sanpera, A. },
title = {Reflections upon separability and distillability},
journal = {J. Mod. Opt},
volume = {49},
number = {8},
pages = {1399},
year  = {2002},
publisher = {Taylor & Francis},


URL = { 
    
        https://doi.org/10.1080/09500340110105975
    
    

},


}

@article{LoopholeB,
  title = {Entanglement witnesses and a loophole problem},
  author = {Skwara, P. and Kampermann, H. and Kleinmann, M. and Bru\ss{}, D.},
  journal = {Phys. Rev. A},
  volume = {76},
  issue = {1},
  pages = {012312},
  numpages = {5},
  year = {2007},
  month = {Jul},
  publisher = {American Physical Society},

  url = {https://link.aps.org/doi/10.1103/PhysRevA.76.012312}
}

@article{Larsson_2014,
url = {https://dx.doi.org/10.1088/1751-8113/47/42/424003},
year = {2014},
month = {oct},
publisher = {IOP Publishing},
volume = {47},
number = {42},
pages = {424003},
author = {Larsson, J. Å.},
title = {Loopholes in Bell inequality tests of local realism},
journal = {Journal of Physics A: Mathematical and Theoretical},
}

@article{Pearle,
  title = {Hidden-Variable Example Based upon Data Rejection},
  author = {Pearle, P. M.},
  journal = {Phys. Rev. D},
  volume = {2},
  issue = {8},
  pages = {1418--1425},
  numpages = {0},
  year = {1970},
  month = {Oct},
  publisher = {American Physical Society},
  url = {https://link.aps.org/doi/10.1103/PhysRevD.2.1418}
}

@article{Santos,
  title={Unreliability of performed tests of Bell's inequality using parametric down-converted photons},
  author={Santos, E.},
  journal={Phys. Lett. A},
  volume={212},
  number={1-2},
  pages={10--14},
  year={1996},
  publisher={Elsevier},
 url = {https://www.sciencedirect.com/science/article/pii/037596019600028X}
}

@article{tele,
  title = {Teleporting an unknown quantum state via dual classical and Einstein-Podolsky-Rosen channels},
  author = {Bennett, C. H. and Brassard, G. and Cr\'epeau, C. and Jozsa, R. and Peres, A. and Wootters, W. K.},
  journal = {Phys. Rev. Lett.},
  volume = {70},
  issue = {13},
  pages = {1895--1899},
  numpages = {0},
  year = {1993},
  month = {Mar},
  publisher = {American Physical Society},
  doi = {10.1103/PhysRevLett.70.1895},
  url = {https://link.aps.org/doi/10.1103/PhysRevLett.70.1895}
}

@article{TT,
url = {https://doi.org/10.1088/1367-2630/17/7/075015},
year = {2015},
month = {jul},
publisher = {IOP Publishing},
volume = {17},
number = {7},
pages = {075015},
author = {Binder, Felix C and Vinjanampathy, Sai and Modi, Kavan and Goold, John},
title = {Quantacell: powerful charging of quantum batteries},
journal = {New Journal of Physics}
}

@article{Non-lin_wit,
  title = {Nonlinear Entanglement Witnesses},
  author = {G\"uhne, Otfried and L\"utkenhaus, Norbert},
  journal = {Phys. Rev. Lett.},
  volume = {96},
  issue = {17},
  pages = {170502},
  numpages = {4},
  year = {2006},
  month = {May},
  publisher = {American Physical Society},
  doi = {10.1103/PhysRevLett.96.170502},
  url = {https://link.aps.org/doi/10.1103/PhysRevLett.96.170502}
}

@article{lin_wit_1,
title = {Separability of mixed states: necessary and sufficient conditions},
journal = {Physics Letters A},
volume = {223},
number = {1},
pages = {1-8},
year = {1996},
issn = {0375-9601},
doi = {https://doi.org/10.1016/S0375-9601(96)00706-2},
url = {https://www.sciencedirect.com/science/article/pii/S0375960196007062},
author = { Horodecki, M. and  Horodecki, P. and  Horodecki, R.}
}

@article{lin_wit_2,
title = {Entanglement detection},
journal = {Physics Reports},
volume = {474},
number = {1},
pages = {1-75},
year = {2009},
issn = {0370-1573},
doi = {https://doi.org/10.1016/j.physrep.2009.02.004},
url = {https://www.sciencedirect.com/science/article/pii/S0370157309000623},
author = { Gühne, O. and Tóth, G.}
}

@article{chaki_2026_20040539,
  author       = {Chaki, P. and
                  Sen, K. and
                  Sen, U.},
  title        = {Dataset of "Effects of the detection loophole on rival entanglement attestation techniques"},
  month        = may,
  year         = {2026},
  publisher    = {Zenodo},
  doi          = {10.5281/zenodo.20040539},
  url          = {https://doi.org/10.5281/zenodo.20040539},
}
\end{document}